\documentclass[a4paper, 12pt]{article}
\usepackage[pdftex]{graphicx}
\usepackage{amsmath, amssymb}

\begin{document}
\begin{center}
\vspace*{5mm}
{\Large \bf
Inequality in resource allocation and population dynamics models
}\\
\vspace{5mm}
{\bf Masahiro Anazawa}%
\footnote[1]{
tel.: +81-22-305-3931; fax: +81-22-305-3901;
e-mail: anazawa@tohtech.ac.jp.
}\\
\vspace{5mm}
 {\it
  Department of Environment and Energy, \\
  Tohoku Institute of Technology, Sendai 982-8577, Japan
 }
\vspace{5mm}
\end{center}

\begin{abstract}
    The Hassell model has been widely used as a general discrete-time
    population dynamics model that describes both contest and scramble
    intraspecific competition through a tunable exponent.  Since the
    two types of competition generally lead to different degrees of
    inequality in the resource distribution among individuals, the
    exponent is expected to be related to this inequality. However,
    among various first-principles derivations of this model, none is
    consistent with this expectation. This paper explores whether a
    Hassell model with an exponent related to inequality in resource
    allocation can be derived from first principles. Indeed, such a
    Hassell model can be derived by assuming random competition for
    resources among the individuals wherein each individual can obtain
    only a fixed amount of resources at a time. Changing the size of
    the resource unit alters the degree of inequality, and the
    exponent changes accordingly. The Beverton--Holt and Ricker models
    can be regarded as special cases of the derived Hassell model. Two
    additional Hassell models are derived under some modified
    assumptions.
\end{abstract}
\vspace{5mm}
\small{
Key Words: 
  population dynamics, 
  Hassell model, 
  first-principles derivation,
  individual, 
  contest competition, 
  scramble competition
}
\vspace{5mm}


\section{Introduction}
\label{sec:introduction}
The population dynamics of seasonally reproducing species with
non-overlapping generations, such as insects, are often described
using discrete-time population models $N_{t+1}=f(N_t)$, which express
the population size at one generation $N_{t+1}$ as a function of the
population size at the previous generation $N_t$. Classic examples
include the Beverton--Holt model \cite{bevhol1957}, Ricker model
\cite{ric1954}, Hassell model \cite{has1975} and
Maynard--Smith--Slatkin model \cite{maysla1973}. However, these models
were originally introduced as phenomenological models at the
population level and were not derived from more fundamental processes
such as the interactions among individuals.  Therefore, the
individual-level mechanisms underlying these models are not very
clear.  However, in recent years, research to derive these population
models from more fundamental processes has progressed
\cite{gerkis2004, brasum2005b, ana2010, ana2018}.  If a population
model can be derived from basic processes, such as resource
competition among individuals, we can better understand the
individual-level processes and phenomena underlying the model and
clarify how the parameters of the population model are related to the
parameters characterising the corresponding individual-level
processes.

\begin{figure}[tb]
\centering
\includegraphics[width=80mm,clip]{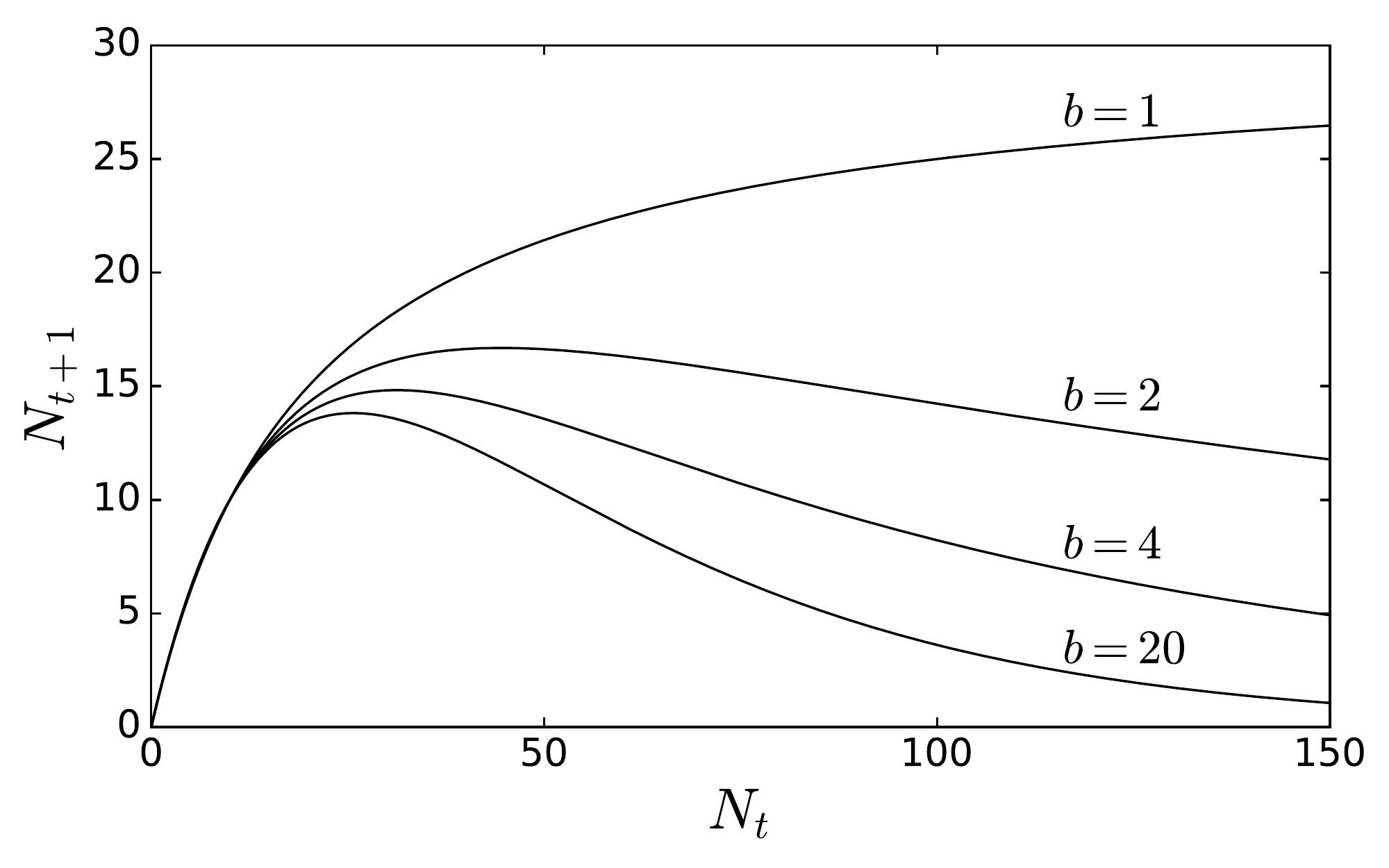}
\parbox{130mm}{\caption{\small %
Reproduction curves of the Hassell model $N_{t+1}=\lambda
N_t/(1+a N_t)^b$ for different values of exponent
$b$. $\lambda=1.5$; $a$ is determined by the condition
$N_{\ast}=10$.
\label{fig:hassell0}
}}
\end{figure}

The Hassell model $N_{t+1}=\lambda N_t/(1+a N_t)^b$ can describe
various types of density dependence by changing the exponent $b$, and
thus it has been widely used as a general population model
\cite{has1975, haslaw1976}.  When $b=1$, the model gives an exact
compensating curve, which monotonically increases towards a maximum as
the population increases (figure~\ref{fig:hassell0}). When $b>1$, the
model gives overcompensating curves, which increase to a maximum and
then decrease with increasing population size. The degree of
overcompensation grows as $b$ increases. The density-dependence types
of exact compensation and overcompensation are often supposed to
reflect different types of intraspecific competition.  Nicholson
\cite{nic1954} emphasised two extreme types of intraspecific
competition, which he called contest and scramble competition. In
ideal contest for resources, a few successful individuals obtain their
full requirements for survival and reproduction, but others obtain
nothing \cite{has1975, vargra1974}. Consequently, when there are many
individuals, the number of successful individuals and their offspring
are almost independent of population size, and the population dynamics
describe an exact compensating curve. In contrast, in ideal scramble
competition, all individuals are assumed to scramble for the resources
and receive an equal share of the resources \cite{has1975,
  vargra1974}.  In a large population, most individuals cannot obtain
sufficient resources for their survival and reproduction, so the
dynamics describe an overcompensating curve. Hassell \cite{has1975}
remarked that in his model, $b=1$ corresponds to ideal contest
competition, the limit $b\to\infty$ corresponds to ideal scramble
competition and $b>1$ corresponds to varying combinations of scramble
and contest. The monopolisation of resources by a few individuals in
ideal contest can be considered as an unequal distribution of
resources \cite{lom2009, lom1988}. Accordingly, exponent $b$ is
expected to be related to the degree of inequality in resource
distribution among the individuals. As $b$ enlarges, the degree of
inequality should decrease. Although the Hassell model was originally
introduced as a phenomenological model at the population level,
several authors have succeeded in deriving it from first
principles. However, as described below, none of these derivations are
consistent with the expected relationship between exponent $b$ and the
inequality in resource allocation.

Existing first-principles derivations of the Hassell model can be
broadly classified into two approaches. The first approach considers
population dynamics in an environment comprising many resource
patches.  De Jong \cite{dej1978, hasmay1985} showed that when the
individuals are distributed negative binomially over the patches,
discrete-time population dynamics are described by a Hassell model.
However, the exponent of her model is $b=k+1$, where $k$ is a clumping
parameter of the negative binomial distribution, and hence it is not
related to the inequality in resource allocation.  In similar ways,
various discrete-time population models \cite{brasum2005b, ana2009,
  ana2010} and interspecific competition models \cite{ana2012int,
  ana2014} have been derived from first principles.  The second
approach assumes a system of differential equations describing
continuous-time dynamics within a year, from which a discrete-time
model for the between-year dynamics is derived.  It is well known that
when a population is subject to a mortality linearly increasing with
the density, the population size after a certain period is described
by a Beverton-Holt model as a function of the original population
size.  Extending this idea, Nedorezov et al. \cite{nednaz1998} derived
a Hassell model by assuming discrete reproduction at the end of the
year with fecundity exponentially decreasing with the mean population
size over the year.  Further, Hassell models have also been derived
from various continuous-time systems incorporating interactions
between species or age classes \cite{thi2003, gerkis2004, eskger2007,
  esk2009, nedvol2008}.  However, similar to de Jong's model, the
exponents of these Hassell models are not related to the inequality
among individuals. In summary, although the Hassell model has been
derived in various ways, no existing derivation has captured the
expected relation between the exponent and the inequality in resource
allocation.

This paper explores whether a Hassell model whose exponent is related
to the inequality can be derived from first principles. Such a Hassell
model can indeed be derived by assuming random competition for
resources among the individuals wherein each individual can obtain
only a fixed amount of resources at a time.  The Beverton-Holt and
Ricker models can be regarded as special cases of the derived Hassell
model, namely, the models for the highest and lowest inequality,
respectively.  Furthermore, two additional Hassell models are derived
when some assumptions are modified.

\section{Derivation of population models}
\label{sec:derivation}

\subsection{Hassell model}
\label{sec:basic}
This section aims to derive a Hassell model whose exponent is related
to the degree of inequality in resource distribution among the
individuals of a population.  The inequality can arise from various
causes, e.g. different competitiveness levels among individuals, but
here it is attributed solely to the randomness in resource
distribution among the individuals.

\begin{figure}[tb]
\centering
\includegraphics[width=80mm,clip]{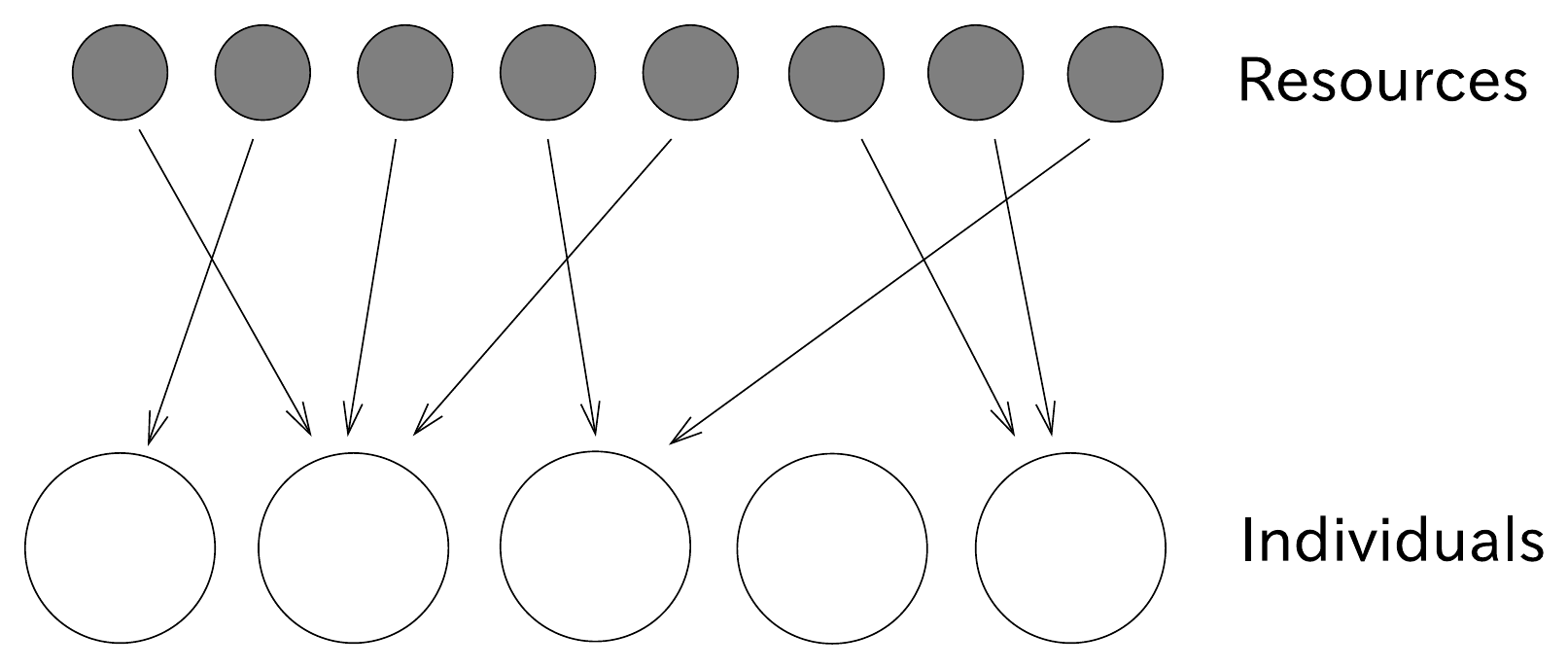}
\parbox{130mm}{\caption{\small %
Competition for limited resources among several individuals.  If
each individual can obtain only a fixed amount of resources at a
time, it follows that the resources are essentially the same as
being divided into many pieces of this fixed size, which are
distributed among the individuals.
\label{fig:basicidea}
}}
\end{figure}

Consider a situation in which $N$ individuals compete for limited
resources. Let us assume that each individual can obtain only a
certain fixed amount $u$ of resources at a time. It follows that the
resources, which may be continuous or discrete, are effectively the
same as being divided into many pieces of this fixed size (resource
units), which are collected by the competing individuals (see
figure.~\ref{fig:basicidea}). When there are $M$ resource units in
total (i.e. the total resource amount is $R=uM$), the competition for
each resource unit is repeated $M$ times.  Assuming that each resource
unit is consumed by a randomly selected individual, the total number
of resource units obtained by a given individual after the competition
follows a binomial distribution. More specifically, the probability
$p_m$ of a given individual obtaining exactly $m$ resource units is
\begin{equation}
  p_m(M)=\left(\frac{1}{N}\right)^m
    \left(1-\frac{1}{N}\right)^{M-m}
    \left({M \atop m} \right),
\label{eq:pm}
\end{equation}
where $m\le M$. The variance of the actual amount of resources
allocated to an individual $um$ is then given by
\begin{equation}
 \mathrm{Var}[um]=uR\left(1-\frac{1}{N}\right)\frac{1}{N}.
\end{equation}
Note that when $R$ is fixed, this variance is proportional to the unit
size $u$. This shows that the inequality among individuals increases
with unit size. Assuming that an individual requires at least $s$
resource units to reproduce, the probability that a given individual
successfully reproduces is
\begin{equation}
  Q(M)=\sum_{m=s}^{M} p_m(M).
\label{eq:QM}
\end{equation}
Letting $f(N)$ denote the expected population at the next generation,
the population model is then written as
\begin{equation}
  f(N)=\lambda N Q(M),
\label{eq:model0}
\end{equation}
where $\lambda$ is the expected number of offspring produced by a
reproductive individual.

Unfortunately, equation~(\ref{eq:model0}) does not give a Hassell
model. We therefore slightly change the assumption such that the total
resource amount $R$ is not a constant but follows an exponential
distribution with the probability density
\begin{equation}
  q(R)=\mathrm{e}^{-R/\bar{R}}/\bar{R},
\label{eq:qR}
\end{equation}
where $\bar{R}$ is the expected value of $R$ (the physical meaning of
this distribution will be discussed later). The total number $M$ of
resource units, i.e. the integer part of $R/u$, follows a geometric
distribution
\begin{equation}
  P_M=(1-\mathrm{e}^{-u/\bar{R}}\,) \, \mathrm{e}^{-M u/\bar{R}}.
\label{eq:PM}
\end{equation}
In this situation, the population model is given by averaging the
right-hand side of equation~(\ref{eq:model0}) with this distribution
as
\begin{equation}
  f(N)=\sum_{M=0}^{\infty} \lambda N Q(M) \cdot P_M.
\label{eq:model1}
\end{equation}
Interestingly, this infinite sum proves to be the following Hassell
model (see appendix~(a) for details):
\begin{equation}
  f(N)=\frac{\lambda N}{\left[1+(\mathrm{e}^{u/\bar{R}}-1)N\right]^s}.
\label{eq:hassell1}
\end{equation}
The exponent $s$ in this model can be written as $s'/u$, where $s'$ is
the actual resource amount required for reproduction.  This shows that
the exponent is inversely proportional to $u$.  As mentioned before,
$u$ is related to the inequality in resource allocation, so the
exponent is indeed related to the inequality. A larger exponent
implies less inequality in resource allocation.  Therefore, we have
succeeded in deriving a Hassell model consistent with the expected
relationship between the exponent and the inequality in resource
allocation.

\subsection{Beverton-Holt  and Ricker models}
\label{sec:special}
Let us consider the largest and smallest inequalities in the derived
Hassell model (\ref{eq:hassell1}). When the unit size $u$ is
considerably large, an individual can reproduce after consuming only
one resource unit, which means $s=1$. In this case, the allocation is
maximally unequal, and the model (\ref{eq:hassell1}) becomes the
following Beverton--Holt
\begin{equation}
  f(N)=\frac{\lambda N}{1+(\mathrm{e}^{u/\bar{R}}-1)N}.
\label{eq:beverton-holt}
\end{equation}

In contrast, the allocation is equalised in the limit $u\to 0$ with
$s'=us$ fixed ($s=s'/u\to\infty$). In this limit, the resources are
equally partitioned among all individuals, and
equation~(\ref{eq:hassell1}) becomes the Ricker model
\begin{equation}
  f(N)=\lambda N \exp(-s'/\bar{R}\cdot N).
\label{eq:ricker}
\end{equation}
Therefore, the Beverton--Holt and Ricker models can both be
interpreted as special cases of the Hassell model (\ref{eq:hassell1})
in which the allocation is maximally unequal and completely equal,
respectively.

\begin{figure}[tb]
\centering
\includegraphics[width=80mm,clip]{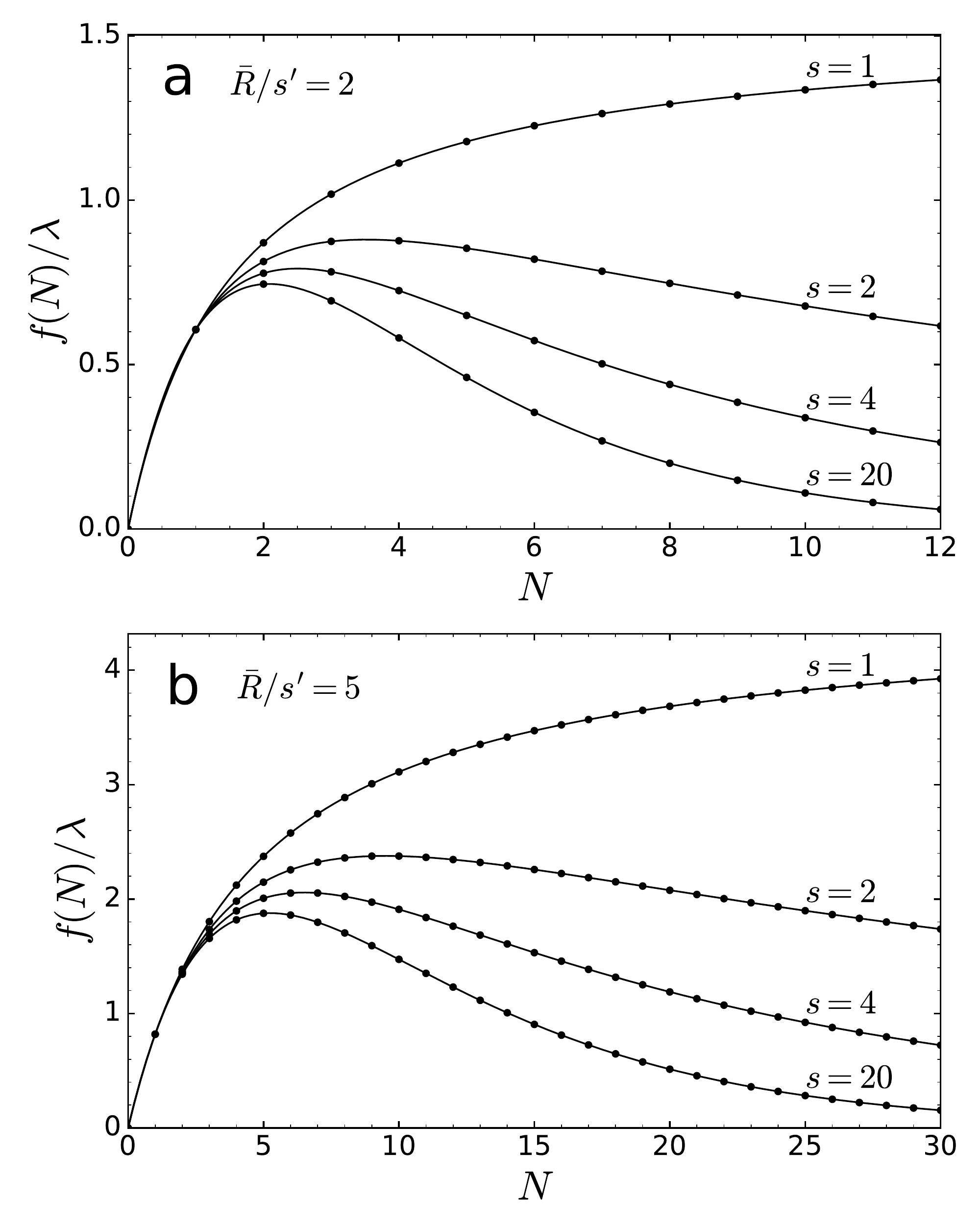}
\parbox{130mm}{\caption{\small %
Reproduction curves of the derived Hassell model
(\ref{eq:hassell1}) for different values of $s$: (a)
$\bar{R}/s'=2$ and (b) $\bar{R}/s'=5$.  As $s$ increases (the
resource unit size $u$ decreases), the degree of inequality in
the allocated resources decreases, and the curves develop higher
degrees of overcompensation.
\label{fig:hassell}
}}
\end{figure}

Figure~\ref{fig:hassell} shows curves of the Hassell model
(\ref{eq:hassell1}) for several values of $s$. As $s$ increases ($u$
decreases), the degree of inequality decreases, increasing the degree
of overcompensation.

\subsection{Interpretation of the derivation}
\label{sec:interpretation}
When deriving the Hassell model (\ref{eq:hassell1}), we assumed that
the total resource amount $R$ follows the exponential distribution
(\ref{eq:qR}).  This assumption needs to be interpreted.  In general,
the following fact is known.  Consider a continuous variable with an
unknown probability distribution. When only the expected value of the
variable is known, the probability distribution that maximises
information entropy turns out to be an exponential distribution
\cite{har2011, harnew2014}. The distribution that maximises
information entropy depends on the prior knowledge we have. For
example, when we know the range of realised values but not the
expected value, the distribution that is both consistent with this
prior knowledge and maximises information entropy is the uniform
distribution whose domain is identical to that range. As suggested by
this example, the distribution that maximises information entropy is
the least-biased prediction of the distribution consistent with the
prior knowledge, which describes the most natural prediction.  In this
interpretation, the Hassell model (\ref{eq:hassell1}) represents the
most natural estimate of the expected number of individuals in the
next generation when only the expected amount of total resources is
known.

\section{Extended models}
\label{sec:extended}
%
\subsection{Non-constant resource threshold}
\label{sec:nc-threshold}
In section~\ref{sec:derivation}, we assumed that all individuals need
the same number $s$ of resource units for reproduction. However, in
real populations, $s$ should vary among the individuals.  The
following discussion considers population models with variable
reproduction requirements.  Interestingly, under certain conditions,
we can obtain Hassell models that include the effect of $s$ variation
among individuals in their exponents.

\begin{figure}[tb]
\centering
\includegraphics[width=80mm,clip]{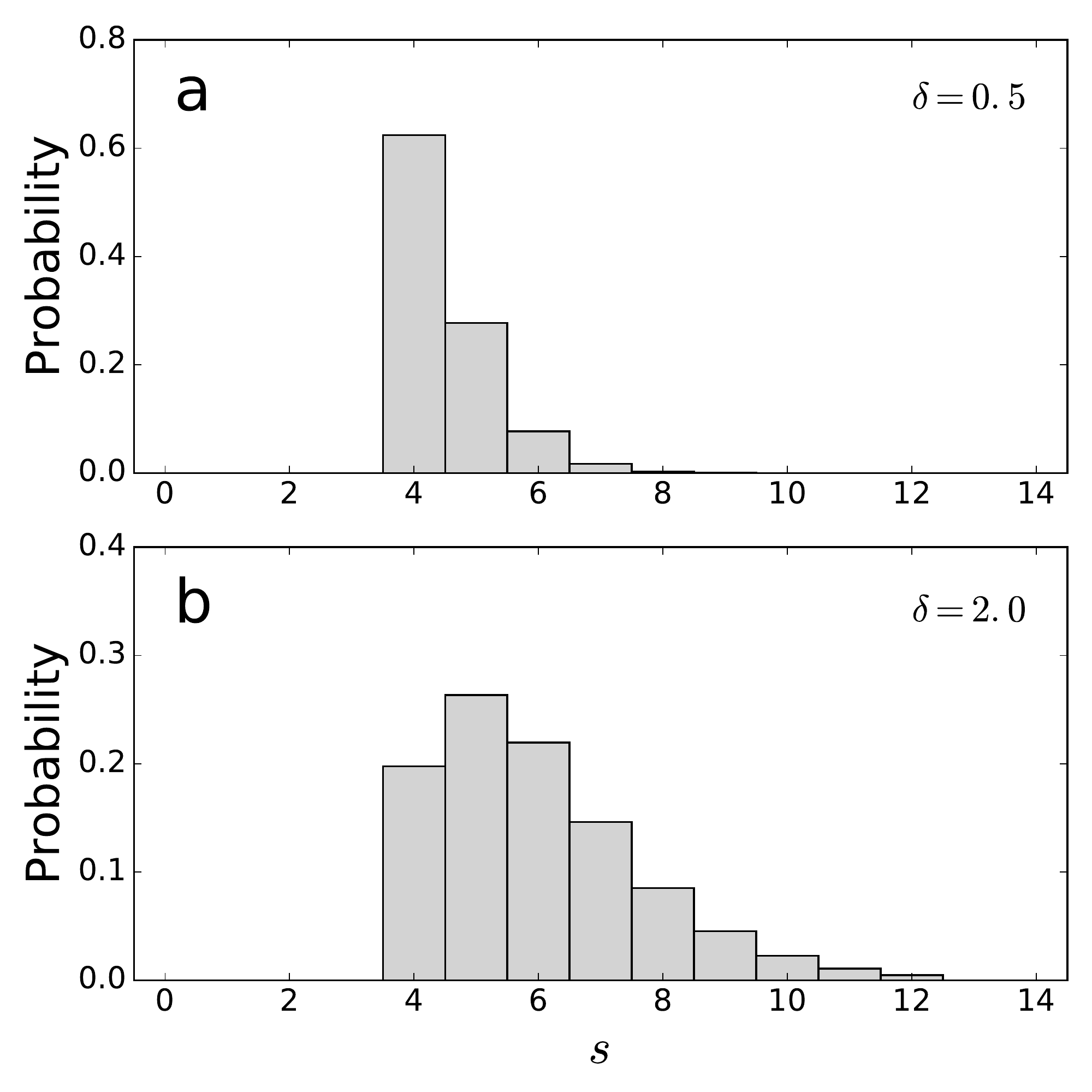}
\parbox{130mm}{\caption{\small %
Examples of the distribution of $s$ defined by equation~(\ref{eq:nbd})
for (a) $\delta=0.5$ and (b) $\delta=2.0$. The variance enlarges
as $\delta$ increases. $k=s_0=4.0$.
\label{fig:nbd}
}}
\end{figure}

When $s$ varies among the individuals, the expected number of individuals
in the next generation can be written as
\begin{equation}
  f(N)=\sum_{s=0}^{\infty} 
  \lambda N \bigl[1+(\mathrm{e}^{u/\bar{R}}-1)N\bigr]^{-s} \, p_s,
\label{eq:sum-s}
\end{equation}
where $p_s$ is the probability of a given individual requiring $s$
resource units for reproduction. As an example, let us consider a case
in which $p_s=0$ for $s$ less than a positive integer $s_0$ and for
$s\ge s_0$, $s-s_0$ follows a negative binomial distribution
\begin{equation}
  p_s=\frac{\Gamma(s-s_0+k)}{\Gamma(k)\Gamma(s-s_0+1)}
    \left(\frac{\delta}{k}\right)^{s-s_0}
    \left(1+\frac{\delta}{k}\right)^{-s+s_0-k},
\label{eq:nbd}
\end{equation}
where $\delta=\bar{s}-s_0\,(>0$), $\bar{s}$ is the mean of $s$ and
$k\, (>0)$ is a shape parameter (see figure.~\ref{fig:nbd}). The
variance of $s$ is $\sigma_s^2=\delta + \delta^2/k$, so it grows with
increasing $\delta$ or decreasing $k$. In the limit $\delta\to 0$, the
variance vanishes, recovering the constant $s$ in
section~\ref{sec:derivation}.  Substituting equation~(\ref{eq:nbd})
into equation~(\ref{eq:sum-s}) and summing the terms gives the
following population model (see appendix~(b) for detail):
\begin{equation}
  f(N)=\frac{\lambda N}{\left[1+(\mathrm{e}^{u/\bar{R}}-1)N\right]^{s_0}}
    \left(\frac{1+(\mathrm{e}^{u/\bar{R}}-1)N}
    {1+(1+\delta/k)(\mathrm{e}^{u/\bar{R}}-1)N}\right)^k.
\label{eq:model3-1}
\end{equation}

Interestingly, after imposing an additional condition $k=s_0$, this
model transforms into a Hassell model
\begin{equation}
  f(N)=\frac{\lambda N}
  {\left[1+(1+\delta/k)(\mathrm{e}^{u/\bar{R}}-1)N\right]^{k}}.
\label{eq:hassell2}
\end{equation}
Even under this condition, the variance of $s$ can be changed freely
by varying $\delta$ (see figure~\ref{fig:nbd}). As evidenced below,
the exponent $k$ of this model includes the effect of varying $s$
among the individuals. From $\sigma_s^2=\delta (1+\delta/k)$ and
$\bar{s}=k+\delta$, we have
\begin{equation}
  k = \frac{\bar{s}}{1+\sigma_s^2/\bar{s}}
      = \frac{\bar{s}'}{u+\sigma_{s'}^2/\bar{s}'},
\label{eq:k}
\end{equation}
\begin{equation}
  \delta = \frac{\sigma_s^2}{1+\sigma_s^2/\bar{s}}
      = \frac{\sigma_{s'}^2/u}{u+\sigma_{s'}^2/\bar{s}'},
\label{eq:delta}
\end{equation}
where $\bar{s}'$ and $\sigma_{s'}^2$ are, respectively, the mean and
variance of $s'=us$, i.e. the actual resource amount required for
reproduction.  Note that the exponent $k$ in equation~(\ref{eq:k})
depends on $\sigma_{s'}^2$ and the unit size $u$. Therefore, it
includes the effects of varying $s'$ along with the inequality in the
allocated resources. Increasing $\sigma_{s'}^2$ decreases the exponent
$k$, thereby reducing the level of overcompensation of the
reproduction curve.  This is probably because a large variation in
$s'$ suppresses an abrupt increase in individuals that fail to
reproduce when the population size increases.

\begin{figure}[tb]
\centering
\includegraphics[width=80mm,clip]{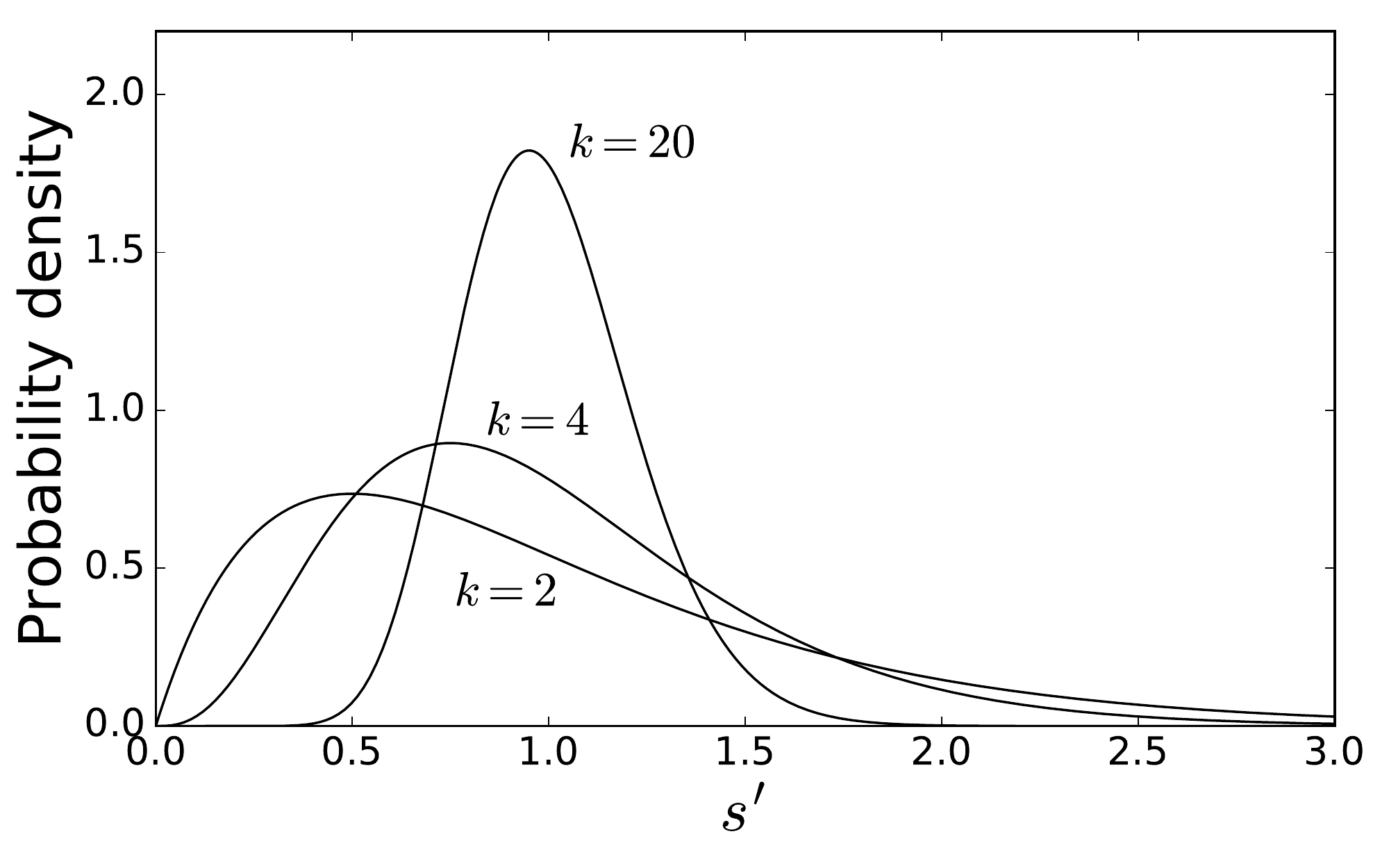}
\parbox{130mm}{\caption{\small %
Examples of the gamma distribution (\ref{eq:gammad}).  The
negative binomial distribution (\ref{eq:nbd}) transforms to this
distribution in the limit $u\to 0$. The variance decreases with
increasing $k$.  $\bar{s}'=1$.
\label{fig:gammad}
}}
\end{figure}

As shown in equation~(\ref{eq:k}), due to the $\sigma_{s'}^2$ term in
the denominator, the exponent of model (\ref{eq:hassell2}) does not
approach infinity in the limit $u\to 0$, unlike Hassell model
(\ref{eq:hassell1}).  Accordingly, even in the limit of equal
distribution of resources, the model remains a Hassell model as
follows:
\begin{equation}
  f(N)=\frac{\lambda N}{\left[1+\bar{s}'/(k\bar{R}) \cdot N\right]^k}.
  \label{eq:hassell3}
\end{equation}
In this case, the exponent $k=\bar{s}'^2/\sigma^2_{s'}$ includes only
the effect of the variation in $s'$.  Note that the same model can be
obtained from model (\ref{eq:model3-1}), i.e. the model before
imposing the condition $k=s_0$, by taking the limit $u\to 0$ with
$\bar{s}'=u \bar{s}$ fixed.
Accordingly, the exponent must be an integer in model
(\ref{eq:hassell2}) but can be a real number in model
(\ref{eq:hassell3}). It is also worth noting that in this limit, the
distribution (\ref{eq:nbd}) transforms into a continuous one,
specifically a gamma distribution with mean $\bar{s}'$, shape
parameter $k$ ($>0$) and probability density
\begin{equation}
  q(s') = \frac{k^k}{\Gamma(k) \bar{s}'}
  \left(\frac{s'}{\bar{s}'}\right)^{k-1} 
  \exp\left(-k\, \frac{s'}{\bar{s}'}\right).
\label{eq:gammad}
\end{equation}
Here, the variance of $s'$ is $\sigma_{s'}^2=\bar{s}'^2/k$, which
decreases as $k$ increases, sharpening the peak in the distribution
(figure~\ref{fig:gammad}).  Thus, equation~(\ref{eq:hassell3}) can be
considered as the population model when $s'$ follows the above gamma
distribution and the resources are distributed equally.

\subsection{Non-constant fecundity}
\label{sec:nc-fecundity}
In section~\ref{sec:derivation}, the fecundity (expected number of
offspring produced by an individual) was assumed constant when the
obtained resource amount exceeded a threshold. However, this
assumption may be overly idealised. In the following discussion, we
relax this assumption and consider population models in which the
fecundity depends on the amount of allocated resources.

The fecundity should gradually increase towards a maximum as the
amount of allocated resources increases. Therefore, we describe the
fecundity by the following function of the number $m$ of the allocated
resource units:
\begin{equation}
  \lambda_m = \begin{cases}
    \, \lambda \left(1-\mathrm{e}^{-\gamma u(m-s+1)}\right) & (m \ge s) \\
    \, 0 & (m < s)
  \end{cases},
\label{eq:fec}
\end{equation}
where $s$ is a positive integer and $\lambda$ and $\gamma$ are
positive parameters. When $m$ is greater than or equal to $s$,
$\lambda_m$ increases towards the maximum $\lambda$ with increasing
$m$.  In the limit $\gamma\to\infty$, equation~(\ref{eq:fec}) reverts
to the case of section \ref{sec:derivation}, with $\lambda_m=\lambda$
for $m\ge s$ and $\lambda_m=0$ for $m<s$. The expected number of
individuals in the next generation can be written as
\begin{equation}
  f(N)= N \sum_{m=s}^{\infty} \lambda_m \, \hat{p}_m,
\label{eq:sum-m}
\end{equation}
where $\hat{p}_m$ is the probability that a given individual obtains
exactly $m$ resource units when $R$ follows the exponential
distribution (\ref{eq:qR}). $\hat{p}_m$ is determined from
\begin{equation}
  \hat{p}_m = \sum_{M=0}^{\infty} p_m(M)\, P_M,
\label{eq:hpm1}
\end{equation}
and a calculation with equations~(\ref{eq:pm}) and (\ref{eq:PM})
yields the following geometric distribution (see appendix~(a)):
\begin{equation}
  \hat{p}_m = 
    \left(\frac{1}{(\mathrm{e}^{u/\bar{R}}-1)N}\right)^m
    \left(1+\frac{1}{(\mathrm{e}^{u/\bar{R}}-1)N}\right)^{-m-1}.
\label{eq:hpm2}
\end{equation}
Substituting equations~(\ref{eq:hpm2}) and (\ref{eq:fec}) into
equation~(\ref{eq:sum-m}), followed by the summation of the terms, the
population model can be expressed as
\begin{equation}
  f(N)=\frac{\lambda N}{\displaystyle
    \bigl[1+(\mathrm{e}^{u/\bar{R}}-1)N\bigr]^{s-1}
    \bigl[1+\xi(\mathrm{e}^{u/\bar{R}}-1)N\bigr]},
\label{eq:model3-2}
\end{equation}
where $\xi=(1-\mathrm{e}^{-\gamma u})^{-1}$.  On comparing this model
with the Hassell model (\ref{eq:hassell1}), we find that $N$ is
replaced with $\xi N \,(>N)$ in one of the $s$ multiplicative factors
of $[1+(\mathrm{e}^{u/\bar{R}}-1)N]$ in the denominator of
equation~(\ref{eq:hassell1}).  Accordingly, this model gives values
smaller than those given by the Hassell model (\ref{eq:hassell1}),
reflecting fecundity (\ref{eq:fec}) lower than that assumed in the
previous model, particularly when $\gamma$ is small.

In the case of highest inequality ($s=1$),
equation~(\ref{eq:model3-2}) becomes the Beverton--Holt model
\begin{equation}
  f(N)=\frac{\lambda N}{1+\xi(\mathrm{e}^{u/\bar{R}}-1)N}.
\end{equation}
Conversely, in the case of equal distribution (in the limit $u\to 0$
with $s'=us$ fixed), equation~(\ref{eq:model3-2}) becomes
\begin{equation}
  f(N)=\lambda N \exp\left(-s'/\bar{R}\cdot N\right)
    \,\frac{1}{1+N/(\gamma\bar{R})}.
\end{equation}
This model has the form of the Ricker model (\ref{eq:ricker})
multiplied by the growth rate of a Beverton--Holt model. Owing to the
multiplicative factor, the values in this model are lower than those
in the Ricker model (\ref{eq:ricker}).

\section{Discussion}
\label{sec:discussion}
The Hassell model was originally intended to describe both contest and
scramble competition by means of a tunable exponent \cite{has1975}.
In this interpretation, the exponent is expected to relate to the
degree of inequality in the resource allocation, but no
first-principles derivation consistent with this expectation has ever
been reported.  In this paper, we have shown that such a Hassell model
can indeed be derived from first principles.  This shows that the
original Hassell model is not only a convenient expressive population
model but also a model with a clear individual-level background.
Changing the size of the resource unit alters the degree of inequality
and the exponent changes accordingly.  In particular, for the cases of
the highest and lowest inequalities, the derived Hassell model becomes
a Beverton--Holt model and a Ricker model, respectively. Therefore,
the three distinct models of Beverton--Holt, Ricker and Hassell can be
understood in a unified way through the size of the resource unit.

As stated in the Introduction section, several first-principles
derivations of the Hassell model have been published in the
literature.  The present derivations do not discredit the earlier
derivations but show that the model can be derived in various
situations. This paper emphasises that only the present derivation of
model (\ref{eq:hassell1}) is consistent with the original
interpretation of the Hassell model \cite{has1975}. In Hassell's
interpretation, $b=1$, the limit $b\to \infty$ and $b>1$ correspond to
contest competition, scramble competition and varying combinations of
scramble and contest, respectively \cite{has1975}.  In the present
derivation, $b=1$ indeed corresponds to the highest inequality
condition of ideal contest. Increasing $b$ reduces the degree of
inequality, and the competition approaches the equally distributed
resource condition of ideal scramble in the limit $b\to\infty$, again
consistent with Hassell's interpretation.  The Hassell models derived
for patchy habitats describe a very different situation \cite{
  brasum2005b, ana2010, dej1978}. In this case, the exponent is
related only to the degree of clumping of the individuals over the
patches, so increasing the exponent does not alter the degree of
inequality or the type of competition at all.  In
section~\ref{sec:extended}, we considered a situation in which the
amount of resources $s'$ required for reproduction varies among
individuals and derived another Hassell model (\ref{eq:hassell2}).
The exponent of this equation is related not only to the inequality in
resource allocation but also to the variance of $s'$.  This result
clearly shows that multiple factors can affect the exponent, providing
new insights into the density effects underlying the Hassell model.

This paper assumed a simple resource distribution model and showed
that the resource unit size affects the resource inequality and thus
the degree of overcompensation of the reproduction curve.  Such a
relationship between the unit size and the resource inequality is
expected not only in this case but also in more general
situations. Even in a more complicated model of resource distribution,
the degree of resource inequality should depend on the size of the
resource unit that an individual can obtain at a time.  Therefore, it
should be a universal property independent of the details of the
distribution of resources that the degree of overcompensation grows
with decreasing unit size.  The inequality in this paper was
attributed only to the random resource distribution among the
individuals. However, other causes of inequality are plausible.  For
example, suppose that every time an individual obtains a resource
unit, its probability of obtaining another resource unit increases. In
this circumstance, the inequality would be higher than assumed herein.
Another source of inequality is the different inherent competitiveness
among the individuals.  The present paper focused only on the
inequality arising from random resource competition in order to reveal
the minimum set of assumptions necessary to derive a Hassell model
from the individual level.

The Maynard--Smith--Slatkin model $N_{t+1}=\lambda N_t/(1+a N_t^b)$ is
a classic population model similar to the Hassell model
\cite{maysla1973}. This model is also expected to describe both
contest and scramble competition by changing the exponent $b$, but
little is known about its individual-level background.  It remains an
open question whether this model can also be derived in such a way
that the exponent is related to resource inequality.  The exponents of
the Hassell models (\ref{eq:hassell1}) and (\ref{eq:hassell2}) are
limited to integers, but Hassell models with real number exponents are
often used phenomenologically.  I attempted to derive real number
exponents but succeeded only in model (\ref{eq:hassell3}), which is
limited to equal resource distributions.  Although this study
considered only single-species population dynamics, the extension to
two-species cases is straightforward when both species have the same
resource unit size, and an interspecific competition model can be
derived.  In general, however, the unit size should depend on the
species.  This situation is not easily extendible from the present
discussion and requires a more careful analysis.  In this case, which
is beyond the scope of this research, the competition between species
of different competition types could be discussed by changing the unit
size and hence the degree of inequality of each species independently.

\vskip1pc




\section*{Acknowledgments}
I am grateful to Takenori Takada, Kazunori Sato, Matthew Holden and
Gabriela Gomes for useful comments and discussions.

\renewcommand{\theequation}{A.\arabic{equation}}
\setcounter{equation}{0} 
\section*{Appendix}

\subsection*{(a) Derivation of equation~(\ref{eq:hassell1}) }
\label{sec:app_a}
The Hassell model (\ref{eq:hassell1}) can be derived from
equation~(\ref{eq:model1}) as follows. As $p_m(M)=0$ for $m>M$,
equation~(\ref{eq:QM}) can be rewritten as
\begin{equation}
  Q(M)=\sum_{m=s}^{\infty} p_m(M),
\label{eq:QM-a}
\end{equation}
and then substituting this equation into equation~(\ref{eq:model1})
gives
\begin{equation}
  f(N)=\lambda N \sum_{M=0}^{\infty} \sum_{m=s}^{\infty} p_m(M) \,P_M.
\end{equation}
Interchanging the order of the two summations, we can rewrite this
equation as
\begin{equation}
  f(N)=\lambda N \sum_{m=s}^{\infty} \hat{p}_m,
\label{eq:fN-a}
\end{equation}
where
\begin{equation}
  \hat{p}_m=\sum_{M=0}^{\infty} p_m(M) \,P_M
\label{eq:hpm-a1}
\end{equation}
represents the probability of a given individual obtaining exactly $m$
resource units when $M$ follows the geometric distribution
(\ref{eq:PM}). Substituting equations~(\ref{eq:pm}) and (\ref{eq:PM})
into equation~(\ref{eq:hpm-a1}) and recalling that $p_m(M)=0$ for
$m>M$, we have
\begin{equation}
  \hat{p}_m=\sum_{M=m}^{\infty} c\, a^M
   \Bigl(\frac{1}{N}\Bigr)^m \Bigl(1- \frac{1}{N}\Bigr)^{M-m}
   \frac{M!}{m!(M-m)!},
\label{eq:hpm-a2}
\end{equation}
where $a=\mathrm{e}^{-u/\bar{R}}$ and $c=1-a$. Relabelling the index
$M$ as $M=M'+m$, we have
\begin{equation}
  \hat{p}_m=\sum_{M'=0}^{\infty} c\, a^{M'+m}
   \Bigl(\frac{1}{N}\Bigr)^m \Bigl(1- \frac{1}{N}\Bigr)^{M'}
   \, \frac{(M'+m)!}{m! M'!}.
\label{eq:hpm-a3}
\end{equation}
Applying the expansion formula
\begin{equation}
  (1-x)^{-k-1}=\sum_{n=0}^{\infty} x^{n}
    \frac{(n+k)!}{k! n!},
\label{eq:exp}
\end{equation}
which holds for $|x|<1$, to the right-hand side of
equation~(\ref{eq:hpm-a3}) with $n=M'$, $k=m$ and $x=a(1-1/N)$, we
obtain
\begin{equation}
  \hat{p}_m=\left(\frac{1}{(a^{-1}-1)N}\right)^m
   \left(1+\frac{1}{(a^{-1}-1)N}\right)^{-m-1}.
\label{eq:hpm-a4}
\end{equation}
Substituting this equation into equation~(\ref{eq:fN-a}) and summing
with respect to $m$, we finally obtain the Hassell model
(\ref{eq:hassell1}).

\subsection*{(b) Derivation of equation~(\ref{eq:model3-1})}
\label{sec:app_b}
Model (\ref{eq:model3-1}) can be simply derived from
equation~(\ref{eq:sum-s}) using a probability-generating function.
Note that equation~(\ref{eq:sum-s}) can be rewritten as
\begin{equation}
  f(N)=\mathrm{E}\left[\lambda N \bigl[1+(\mathrm{e}^{u/\bar{R}}-1)N 
    \bigr]^{-s}\right],
\end{equation}
where $\mathrm{E}[A(s)]$ represents the expected value of a function
$A(s)$ of a random variable $s$. This equation can be rewritten as

\begin{equation}
  f(N)=\lambda N\, G\left(\bigl[1+(\mathrm{e}^{u/\bar{R}}-1)N 
    \bigr]^{-1}\right),
\label{eq:model-moment}
\end{equation}
where $G(z)\equiv\mathrm{E}[z^s]$ is the probability-generating
function of $s$. As equation~(\ref{eq:nbd}) shows, $s-s_0$ follows the
negative binomial distribution with mean $\delta$ and shape parameter
$k$, and its generating function is known to be
\begin{equation}
  \mathrm{E}\left[z^{s-s_0} \right]
  = \left( 1+(1-z)\delta/k \right)^{-k}.
\label{eq:mge2}
\end{equation}
The generating function of $s$ is written as
$G(z) = z^{s_0} \mathrm{E}\left[z^{s-s_0}\right]$,
and substituting equation~(\ref{eq:mge2}) into the right-hand side of
this equation yields
\begin{equation}
  G(z)=z^{s_0}\left(1+(1-z)\delta/k\right)^{-k}.
\label{eq:mge}
\end{equation}
Applying this function to equation~(\ref{eq:model-moment}) finally
gives model (\ref{eq:model3-1}).


\end{document}